\documentclass[runningheads]{llncs}
\pdfoutput=1
\usepackage{graphicx}
\usepackage{float}
\usepackage{citesort}
\begin{document}
\title{Nothing But Net: Invading Android User Privacy Using Only Network Access Patterns}
\titlerunning{Nothing But Net}
%
%

\author
{
Mikhail Andreev \inst{1} 
\and
Avi Klausner \inst{1}
\and
Trishita Tiwari \inst{1}
\and
Ari Trachtenberg \inst{1}
\and
Arkady Yerukhimovich \inst{2}
}
\authorrunning{M. Andreev et al.}
%
\institute{Boston University, Department of Electrical and Computer Engineering, Boston, MA 02215, USA \\
\email{\{mikh,trtiwari,trachten\}@bu.edu}, 
\email{aviclaws@gmail.com}
\and
MIT Lincoln Laboratory, Lexington, MA 02421, USA\\
\email{arkady@ll.mit.edu}
}
\maketitle              
\let\thefootnote\relax\footnote{DISTRIBUTION STATEMENT A. Approved for public release: distribution unlimited. Arkady Yerukhimovich's work supported by the Defense Advanced Research Projects Agency under Air Force Contract No. FA8702-15-D-0001. Any opinions, findings, conclusions or recommendations expressed in this material are those of the author(s) and do not necessarily reflect the views of the Defense Advanced Research Projects Agency.}

\begin{abstract}
  We evaluate the power of simple networks side-channels to violate user privacy on Android devices.
  Specifically, we show that, using blackbox network metadata alone (i.e., traffic statistics such as transmission time and size of packets)
  it is possible to infer several elements of a user's location and also identify their web browsing history (i.e, which sites they visited).
  We do this with relatively simple learning and classification methods and basic network statistics.
  For most Android phones currently on the market, such process-level traffic statistics are available for any running process, without any permissions control and at fine-grained details, although, as we demonstrate, even
  device-level statistics are sufficient for some of our attacks.  In effect, it may be possible
  for any application running on these phones to identify privacy-revealing elements
  of a user's location, for example, correlating travel 
  with places of worship, point-of-care medical establishments, or political activity.
\end{abstract}


\section{Introduction}
\label{sec:back}

Increasingly, people are using smartphones for essential tasks in their everyday lives such as navigation, communication, scheduling, banking, dating, and the like.  To facilitate this transformation, users gladly install applications (\emph{apps}) and carry them around, well within their personal space and everywhere they go.  Simultaneous with this increased utility is an invariable loss of privacy.

More precisely, modern smartphones are equipped with a wide variety of sensors, ranging from GPS, triaxial accelerometers, gyroscopes, and magnetometers to stereo-microphones and cameras; smartphones are also equipped with software ``sensors'' monitoring data usage, orientation and the like.  In all, these phones collect vast amounts of private information about their users.  Indeed, many app developers use such private data collection as their primary revenue stream, ``freely'' offering the apps in a Faustian exchange for sensor data, which may be used or resold to advertisers~\cite{leontiadis2012don,freeforall,patsakis2015analysis}.  More general privacy concerns for smartphones abound in the literature~\cite{spensky2016,felt2012android,grace2012unsafe}.

In an attempt to alleviate these privacy concerns, the makers of smartphones have provided mechanisms for users to control what sensors and personal data an application may access.  For example, Google's Android requires that applications request permissions to access sensitive data or sensors, either at install time (``Normal'' permissions) or at run time (``Dangerous'' permissions).  In principle, this allows a user to see what information an application wishes to use, and to deny access if the user feels that such access is not warranted.

The reality, however, is that not all sensors and data types require permissions, meaning that some information not deemed sensitive by Google can be accessed by any application.  Furthermore, this allowed information can introduce side-channels that, when combined with other seemingly non-sensitive information, can be used to deduce private details.  For example, Narain et al.~\cite{narain2016inferring} show that using gyroscope, accelerometer, and magnetometer information, all accessible with no permissions on later-model phones, an app can deduce a user's location with fine accuracy (which is otherwise considered sensitive and permissions-protected).  This highlights a major shortcoming of the permissions model: users (and developers) are unlikely to fully comprehend the implications of such side channels, and thus will not know how they are compromising their own privacy by approving permissions requests.

In this paper we investigate how one such side channel can be used to compromise aspects of user privacy on Android devices.  Specifically, we consider the \emph{network side-channel}, which consists of information such as the sizes and timings of packets sent over the network by the Android device.  We note that this data is accessible to any application running on most Android devices thus giving any adversarial app access to this information.  We show that this side-channel is quite powerful in revealing different types of private information about the user through the use of an experimental attack that relies exclusively on network side channel data that is collected, without access to any physical sensors or operating system permissions. Our first attack learns features of a user's physical location using simple statistical analysis, and our second attack learns a user's web browsing habits using simple machine learning. Lastly, we would like to stress that this paper is not about improving the state-of-the-art performance of the specific side-channel attacks; with some solid engineering polish and more sophisticated tools, we are confident that the results can be sharpened, and possibly even significantly so. The aim of the work is to highlight the ease with which some private information is accessible through the network side-channel, despite the current mitigations. Our point is that just about \emph{any} app developer, with only the most basic machine learning skills, can, today, mine the private information of the app's user base. 

\subsection{Results}
Our main results are as follows.

\medskip\noindent
For the case of location privacy, we show that significant elements of a user's location and movement may be inferred from the network side-channel (without requiring any permissions) while another location-privileged app is running.  Specifically, with access to only the timing and size of packets transmitted through Google Maps, we can learn:
\begin{itemize}
	\item whether or not the user is moving,
	\item the user's approximate rate of motion,
	\item some elements of a user's surroundings (e.g. in a city, in a remote location, ...) and,
	\item whether the user traversed certain specific paths.
\end{itemize}

\medskip\noindent
For the case of web browsing, we show that we can differentiate between many of the websites visited by the user, by observing only device level network usage data (i.e., across all applications on the phone).  Specifically, with access to only the aggregate timing and size of packets transmitted by the applications on the device, we can learn:
\begin{itemize}
	\item which of two websites a user visits with greater than 90\% certainty, 
	\item whether a user visits one of 35 common, ``distinctive'', websites with 75\% certainty, and
	\item which of 128 popular websites a user visits with certainty greater than 16\%.
\end{itemize}
These latter results are robust to some standard network traffic analysis countermeasures, such as random session and packet padding.

We note that the location results are valid for most Android phones on the market today, whereas the web browsing results are valid on all Android phones.  In both cases, our results rely on well-used machine learning techniques, and their simplicity and robustness are thus surprising.

\subsection{Paper Organization}
The remainder of this paper is organized as follows.  In Section~\ref{sec:network}, we describe the network side-channel and what types of applications may be vulnerable to its use, followed, in Section~\ref{sec:rel}, with a review of some related work.  Then, in Sections~\ref{sec:re} and~\ref{sec:browse}, we describe the experimental methodology and results for the attacks against location and browsing privacy. Finally, we conclude in Section~\ref{sec:conclusion}.

\section{The Network Side-Channel}\label{sec:network}
The network side-channel is a source of metadata about packets that are transmitted by another
process.  Such metadata typically include precise times when a process transmits and receives packets, and the sizes of these packets.  We assume
that the packets themselves are encrypted and, though the target address of the packets may be generally accessible in plaintext 
(e.g., for routing), it is not specifically accessible for our purposes.

In general, despite the inaccessible contents of transmitted packets, the network side-channel is an excellent source of information for applications that:
\begin{itemize}
	\item Have quality-of-service constraints, and
	\item Are bandwidth-sensitive.
\end{itemize}

One well-studied example of an application that matches these properties is Voice Over IP (VoIP), as described further in Section~\ref{sec:rel}, where voice packets must be routed
to their destination within certain constraints on jitter and latency, or else the received speech sounds unnatural (or possibly unintelligible).  The constrained nature of the communication channels through which VoIP data typically travels necessitates the use of content-based variable-rate coding schemes, so that the very timing of packet transmissions correlates with the speech being transmitted~\cite{VoIP1_Attack,VoIP2_Attack}.

In our case, both of our victim applications, Google Maps and a web browser, also satisfy these two properties.  For Google Maps, the second condition
(bandwidth-sensitivity) derives from the sheer
size of the world map on which Google Maps is based; it is not feasible to store copies of this map on all devices.  Indeed, Google provides overlays that provide dynamic content (from ads or local points of interest) that needs to be provided on-the-fly to the phone.  Therefore, Google provides map contents (``tiles'') piecemeal to a user, depending on the user's movements.  The first condition for network side-channel sensitivity (quality-of-service constraints) is evident from the real-time usage of such mapping services.  When a user is navigating with Google Maps, map information must be received in a timely fashion as the user moves.

For web browsing, the different sizes and locations of page elements on a website translate into variance in the bandwidth required to download a specific web page.  Since users are typically averse to delays in browsing times~\cite{nah2004study}, it is necessary to load webpages reasonably quickly, thereby satisfying the first constraint.

\subsection{Threat} 
Surprisingly enough, for many Android devices on the market, the Google\\\texttt{TrafficStats~API} provides fine-grained network side-channel information on \emph{any running process} without requiring \emph{any} permissions; for these devices it is possible to learn a variety of facets of a user's location and web browsing habits from any running app. 
Indeed, the typical user of a smartphone cannot be expected to understand the inner workings of smartphone applications.  It is thus extremely easy to disguise an application as a game or some other simple utilitarian task, while in the background inferring and even broadcasting sensitive information, such as the user's location or browsing history. With such unpermissioned malware running, anyone receiving the broadcasts could obtain the web browsing history of the smartphone user or elements of the user's location (as long as the user is using Google Maps).  In essence, the capability to infer a user's location and obtain a user's web browsing history by analyzing the network traffic of applications undermines the permission request protocol and potentially compromises user privacy~\cite{loc3_Attack,loc4_Attack}.

\subsection{Android versions}
\label{sec:upSoft}
Recently, Google has taken steps toward further protecting its users' sensitive information from such side-channels.  As of Android Nougat (version 7.0, API 24), Google is phasing out some of the functionality of the \texttt{TrafficStats} class, and replacing it with \texttt{NetworkStatsManager}, which requires a permission in order to obtain network traffic about other applications~\cite{TrafficStatsPhaseOut}. 
Unfortunately, this only helps mitigate, but does not necessarily eliminate, the threat:

\begin{itemize}
	\item The TrafficStats class can still be used to acquire device-level network data (i.e., the total network data of all running applications on the device). This aggregate network data is sufficient for our attacks, with only a modest reduction in effectiveness.
	\item The majority of users have not and will not upgrade their phones to Nougat (or later versions) for some time (if ever)~\cite{androidMarketShare}.
	\item It is not intuitive to either a typical user or a developer that allowing an application to access network traffic data could possibly leak one's location.  Many users may blithely grant network stats permissions to a malicious application disguised as, say, a network usage monitoring tool.
\end{itemize}


\section{Related work} \label{sec:rel}
Side-channel attacks aim to reveal sensitive information from the implementational byproducts of otherwise secure functionality.  One of the classic examples is Kocher's attack on various cryptosystems~\cite{kocher1996timing}, wherein a fine-grained statistical analysis of the amount of \textit{time} required to perform cryptographic operations revealed the secret key used in the operations.  This attack was extended by Genkin, Shamir and Tromer~\cite{genkin2014rsa} to crack the secret key of a 4096-bit RSA cryptosystem using acoustic emanations typically from voltage regulation circuits on a laptop.  Other side-channel attacks make use of analysis of power usage, faults, or caches~\cite{Boneh1997,osvik2006cache,Kocher2011}.

The \textit{network} side-channel has also been a great source for information leakage.  Encrypted VoIP services like Skype leak information about the language spoken and speaker through the timing (and, in some cases, sizes) of their IP packets~\cite{VoIP1_Attack,VoIP2_Attack}.
More advanced models in~\cite{VoIP4_attack,VoIP5_attack} have been able to recreate the actual audio transcript of what is
being said over an encrypted VoIP connection by measuring the same networking statistics.

\subsection{Location privacy}
The closest work in the current literature may be that of Narain et al~\cite{narain2016inferring}, who use a phone's gyroscope, accelerometer, and magnetometer to infer routes and locations with reasonable accuracy; from such data, their results can provide, for example, a list of 10 routes that include an actually traveled route with greater than 50\% probability.  In a similar vein, the taxonomy in~\cite{DBLP:journals/corr/SpreitzerMKM16} highlights a number of other sensor-based side-channel attacks on location, including using accelerometer readings to infer mode of transportation~\cite{hemminki2013accelerometer}, utilizing barometer and elevation to infer driving routes~\cite{ho2015pressure}, using speaker status information~\cite{zhou2013identity} and observed power consumption~\cite{michalevsky2015powerspy}.  Our work differs from these in looking \emph{exclusively} at network traffic, without requiring any sensor data, and yielding reasonable results with very simple machine learning and classification algorithms.

\subsection{Web browsing privacy}
Traffic analysis attacks on web browsing are not new, dating back at least to Heyning Cheng and Ron Avnur's paper in 1998~\cite{1998}.  That work uses
the size of packets, their arrival times, and plaintext metadata to determine the specific page being viewed by a user on an identified host.
Later work focuses on more specific cases where the host information is not available, and all that is given is the packet size and timing.  Those works distinguish themselves in their analyses. As the traffic analysis became more sophisticated, the task of parsing the raw data was given to different machine learning algorithms, with Herrmann~\cite{multi_bayes} proposing the use of a Multinomial Naive-Bayes model with frequency transformations and Panchenko~\cite{onion} utilizing a Support Vector Machine. Combining several such work together, Dyer used these methods to dismantle common countermeasures proposed against this side-channel attack~\cite{peekaboo}.  Other work (e.g., ~\cite{GKB10,GBKS12}) focuses on remote traffic analysis attacks that leverage bandwidth leakage at routers to identify browsing habits of remote hosts and even to launch deanonymization attacks on Tor users and relays~\cite{MKJCB11}.  Finally, the work most similar to ours is that of Spreitzer et al.~\cite{data_usage_statistics}, wherein the authors show extended results of experiments similar to ours for identifying visited websites based on the network side-channel on Android.  Their results use a different set of techniques and appear to require the use of the \texttt{TrafficStats} class to access the data sent by individual processes, whereas our attacks use only device-level network statistics; with the newest version of Android, only device-level data is freely available.   As such, our work shows the attacks presented in previous publications are still effective using device-level, coarser grained network traffic, and that our attacks are robust to certain (but not all) standard countermeasures.


\section{Attacking Location Privacy} 
\label{sec:re}
In this section, we present our attacks on location privacy using the network side channel. Location is a particularly sensitive piece of information since, for example, with regular knowledge of a user's location it is possible to infer other sensitive and private information (e.g. medical conditions, political or religious connections)~\cite{loc3_Attack}, a suitable time to rob the user's home, or a means for stalking the user~\cite{loc4_Attack}.  It is thus with good reason that Android applications wishing to have access to a user's location must typically explicitly request permissions for this information.

We do not claim to provide complete fine-grained location tracking from network patterns alone - this task would be quite ambitious, given that even the Global Positioning System (Standard Positioning Service) guarantees only a global average of 7.8m of accuracy (95\% of the time) with the use of a constellation of 24 satellites and a dedicated receiver~\cite{gps08}.

Instead, we show that we are able to discern, with varying accuracy, a number of potentially sensitive location features using only a black-box analysis of Google Maps network usage patterns via the TrafficStats API. These
features include:
\begin{itemize}
\item{motion} - detecting when a phone is in motion,
\item{speed} - determining how fast the phone is moving,
\item{coarse location} - inferring some coarse sense of where the
phone might be (e.g. in a city, in a remote location, ...), and
\item{path traversal} - detecting when the phone traverses a specified path.
\end{itemize}

Putting these together, despite their individual issues of accuracy, will allow us to construct a reasonably good understanding of the phone's location at a given time.

\subsection{Key Tools} 
\label{sec:tt}
Before we describe the testing methodology, we would like to elaborate on the key tools that we used for the experiments.

\subsubsection{Google Maps}
\begin{figure}[h]
	\centering
   \includegraphics[scale=0.5]{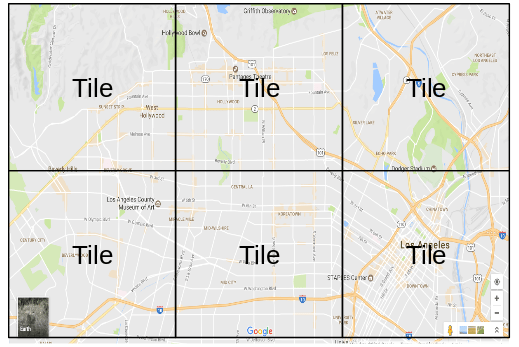}
   \caption{Tiles on Google Maps}
   \label{fig:tiles}
\end{figure}

Google Maps is a real time navigation app provided by Google on Android. The app consists of a screen that depicts a map of the user's location, and as the user moves around, the map is updated with the user's new location. The map itself is composed of a series of tiles, as depicted in Figure \ref{fig:tiles}. Due to the Quality-of-Service constraints of the app (discussed in Section~\ref{sec:network}), updates to the user's map need to happen in real time by downloading new tiles as the user moves around. This process of updating the map view in real time leaks information about the user's location, as one could potentially correlate the rate at which these tiles are downloaded with the user's location information. 

\subsubsection{TrafficStats}
TrafficStats is the Application Programming Interface (API) that allows programs to tap into the network side channel that is inherent to Google Maps. It is provided by the Android Framework and allows~\emph{any} app to monitor certain network statistics, including:
\begin{itemize}
  \item \textbf{App level statistics}: Provide the number of bytes sent/received by a given app since device boot.
  \item \textbf{Device level statistics}: Provide the number of bytes sent/received by the entire device since device boot. 
\end{itemize}

For the most recent versions of Android, only device level statistics are supported, and thus the majority of our tests are focused on extracting location sensitive information from device level traffic (rather than the finer-grained app level traffic).  Indeed, we demonstrate meaningful inferences about a user's location from these device level statistics as long as there is not too much network noise from other applications.  We believe that this low noise assumption is reasonable because many people use Google Maps in isolation while navigating.

\subsubsection{GeoFix}
We conducted two sets of experiments:  one within the Android emulator, and another on an actual device carried around by one of the authors.  Within the emulator, location information was spoofed through \emph{GeoFix}, a framework provided by the Android emulator console.  To utilize this framework, we initiated a socket connection to the emulator (port 5554), and provided a \verb|geo fix| command:
\[
\verb|geo fix <longitude value> <latitude value>|
\]
This allowed us to set the ``location'' of the emulator and move it around the world as needed.  In our experiments, we run this command in a loop, each time incrementing the coordinates slightly to simulate motion in a chosen direction.

\subsection{Experimental Setup}
\label{ExperimentalSetup}
In order to use the TrafficStats API, we created a simple service that runs in the background on an Android device or emulator, logs all app/device level traffic at a 0.5 millisecond granularity, and sends it to a remote server. We do this while simultaneously running Google Maps in the foreground and simulating motion (thereby forcing Google Maps to change the view on the app). Although Google Maps was the only active \emph{foreground} application, there was network noise from other standard background applications (and GeoFix).

We used this setup to conduct four experiments:
\begin{itemize}
\item Detecting when a phone is in motion;
\item Determining how fast the phone is moving;
\item Inferring whether the phone is in a rural vs. urban area; and,
\item Detecting when the phone traverses a specific path.
\end{itemize}

These experiments were conducted at large scale on network data collected from an Android emulator, and then, in some cases, verified on a smaller scale through network data collected from a real device. 
For emulator experiments, we simulated motion with a simple python script that utilized GeoFix to incrementally change the emulator's GPS coordinates. For device experiments, motion was produced by physically driving in the streets of Boston.  Further details of each experiment will be described in the sections that follow. 

\subsection{Data Sets}
\label{sec:Data Set}
We next describe the dataset consisting of device level traffic collected from an unpermissioned background service on a Nexus 6 API 23 emulator running Google Maps and ``traversing'' various paths through GeoFix-based location emulation. Our dataset consists of 432 paths, with paths differing along three elements:
\begin{itemize}
\item \textbf{Location:} the physical location of the path;
\item \textbf{Direction:} the direction in which the path extends (e.g., North, South, East, West or some combination thereof);
\item \textbf{Step interval:} the time delay between subsequent steps on the path (corresponding to the speed of traversal).
\end{itemize}

\paragraph{Location.}
Our paths are distributed among twelve locations equally divided between urban and rural environments, as described in Table~\ref{table:locations}.

\begin{table}
\setlength\tabcolsep{1em}
	\begin{center}
	\begin{tabular}{| c | c | c | c |}
    \hline
	\textbf{Longitude} & \textbf{Latitude} & \textbf{Rural/Urban} & \textbf{Locale} \\
    \hline
	13.75 & 51.05 & Urban & Dresden, Germany \\
	-71.11 & 42.35 & Urban & Boston, USA \\
	77.10 & 28.70 & Urban & New Delhi, India \\
	37.62 & 55.76 & Urban & Moscow, Russia \\
	-105.14 & 39.76 & Urban & Denver, USA \\
	77.35 & 12.95 & Urban & Bangalore, India \\
	13.18 & 13.23 & Rural & Rural Chad \\
	9.53 & 18.86 & Rural & Rural Niger \\
	77.77 & 23.82 & Rural & Rural India \\
	-105.78 & 43.25 & Rural & Rural Wyoming \\
	-109.72 & 48.39 & Rural & Rural Montana \\
	104.15 & 66.53 & Rural & Rural Russia \\
    \hline
	\end{tabular}
\end{center}
\caption{Locations of paths in our dataset.~\label{table:locations}}
\end{table}

\begin{figure}
\centering
   \includegraphics[scale=1]{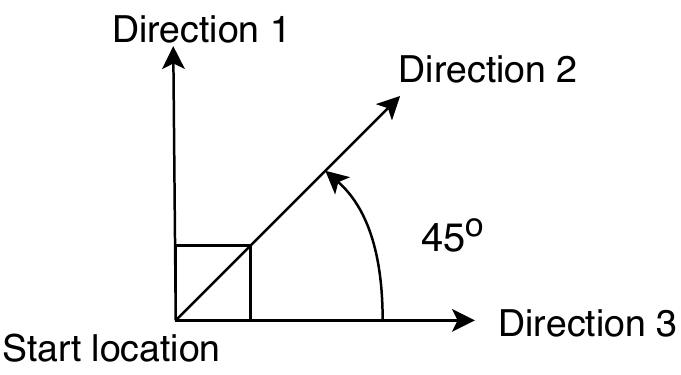}
   \caption{Paths of different directions taken for each location}
   \label{fig:diffDirections}
\end{figure}

\paragraph*{Direction.}
For each of the twelve locations above, we considered three specific directions, as shown in Figure \ref{fig:diffDirections}.

\paragraph{Step interval.}
For each combination of location and direction, we also considered twelve different delay periods between path steps.  More precisely, each path consisted of ten steps, taken at regular time intervals, which varied among twelve different values.  Due to geometric considerations (such as the curvature of the Earth), paths at different locations or directions corresponded to different land speeds, even if their inter-step time intervals were the same.

\paragraph{Summary}
In total, the number of paths we evaluated was
\[
\mbox{\# places} * \mbox{\# directions} * \mbox{\# step intervals} = 12*3*12 = 432.
\]
Within the emulator, each path was attempted eight times to produce eight datasets in total.

\subsection{Experiments and Results}
\subsubsection{Motion}
\label{sec:t_mot}
Our first experiment concerned the ability to detect motion from the network traffic side-channel alone.  To test this, we held the emulator still for fifteen seconds before each path traversal.  In each case, our unpermissioned background service collected network traffic on the emulator while Google Maps was running in the foreground.

\begin{figure}
\centering
   \includegraphics[scale=0.6]{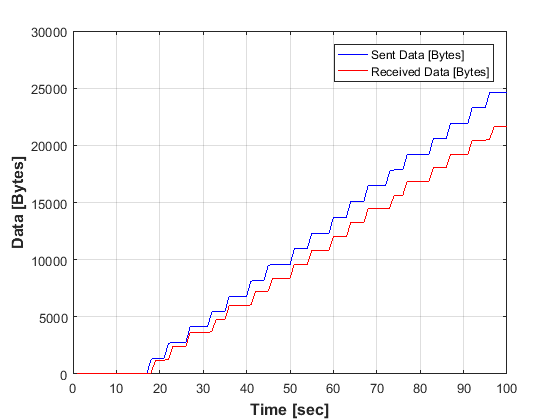}
   \caption{Application-level network traffic statistics of an emulator in eventual motion.}
   \label{fig:motionDetectionApp}
\end{figure}

\begin{figure}
\centering
   \includegraphics[scale=0.6]{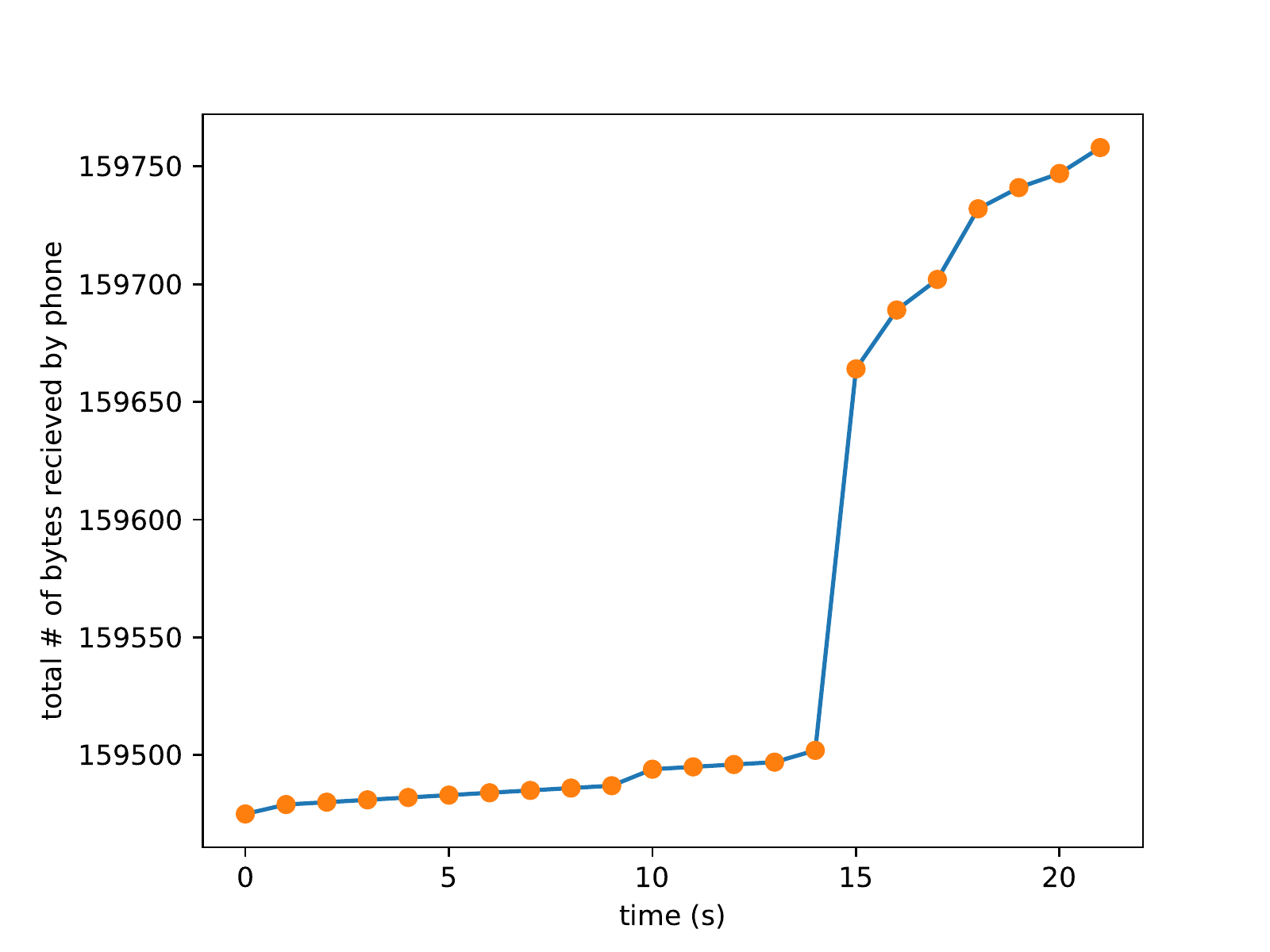}
   \caption{Device-level network traffic statistics of an emulator in eventual motion.}
   \label{fig:motionDetectionDevice}
\end{figure}

Our results show that motion is readily detectable both from application-level network traffic data (Figure~\ref{fig:motionDetectionApp}) and device-level network traffic data (Figure~\ref{fig:motionDetectionDevice}).  In the former case, Google Maps downloads no data until movement starts, and that data thus show a crisp delineation of motion.  In the latter case, there is some background network noise, but the onset of motion is still quite clear at the fifteen second mark.

\subsubsection{Speed}
\label{sec:t_dis}

Our second set of motion experiments attempted to determine whether the \emph{speed} of motion can be inferred from network byte inter-arrival times while Google Maps was in the foreground. Intuitively, the faster the phone moves, the more frequently Google Maps will have to download new tile data. As a result, we expect an inverse relationship between the speed of the device and the time interval between packet arrivals. 

\begin{figure}
\centering
   \includegraphics[scale=0.55]{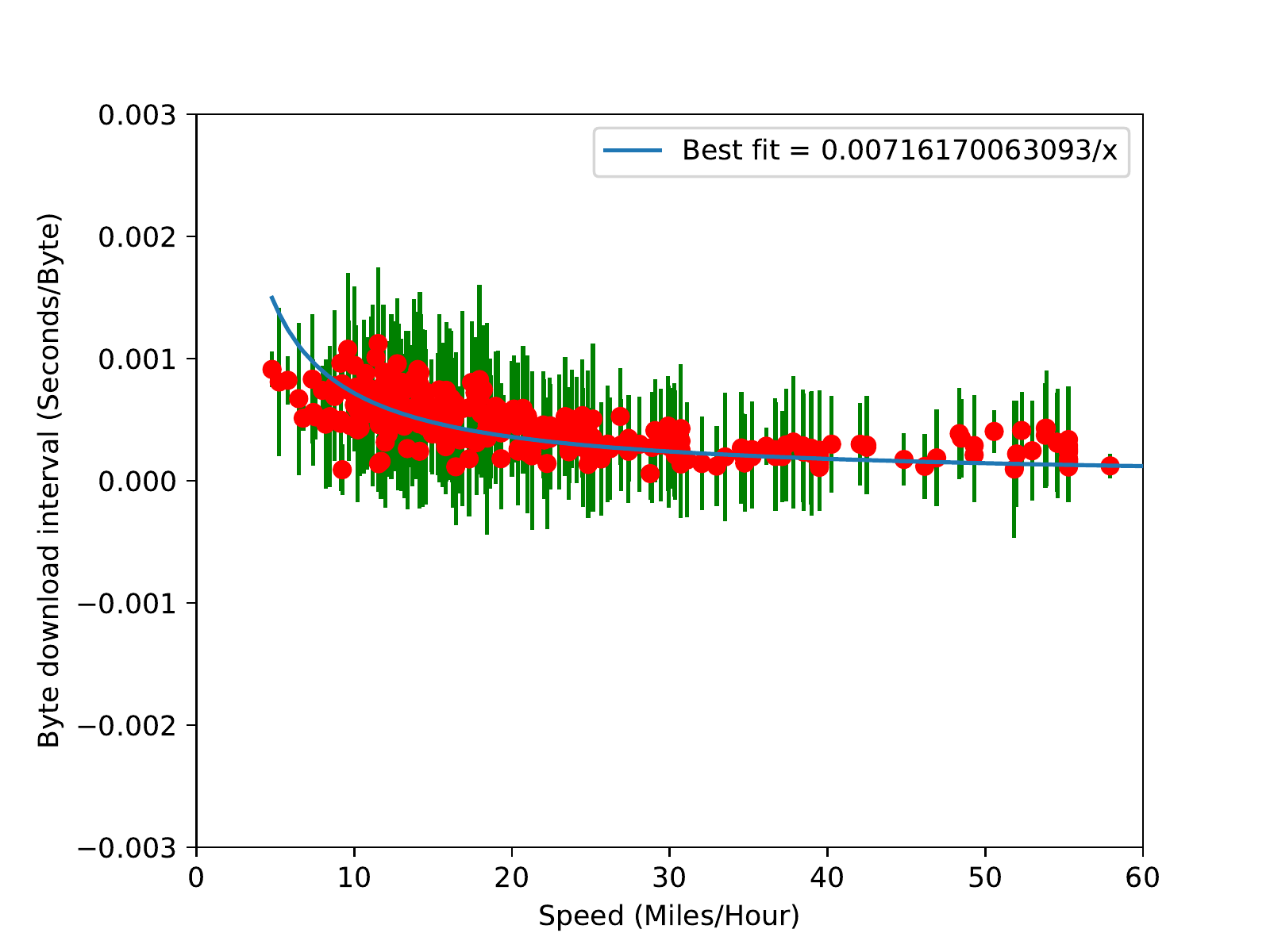}
   \caption{Device-level byte arrival interval vs speed, showing an inverse relationship.}
   \label{fig:speedMotionDevice}
\end{figure}

Since each of our paths corresponded to a different speed (as discussed in Section~\ref{sec:Data Set}), we had a fairly diverse collection of data points for comparing byte inter-arrival rates (in seconds / byte) versus emulated speed (miles / hour).  For device-level data, the result is shown in Figure~\ref{fig:speedMotionDevice}, confirms an inverse relationship, although with significant noise due to other network traffic; indeed, each point on the graph represents an average over eight trials, with error bars representing one standard deviation after removing 10\% of the data point outliers.  For this data, the best fit line is given by:
\[
	y = \frac{0.0071617}{x},
\]
Where the variance on the coefficient is $2.15037546 * 10^{-8}$, suggesting a very good fit.

\begin{figure}
\centering
   \includegraphics[scale=0.55]{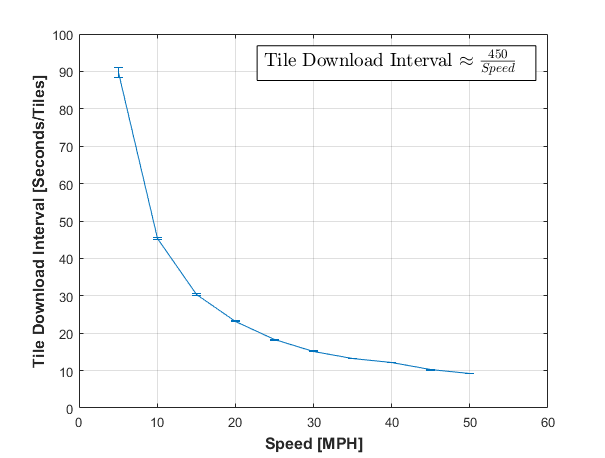}
   \caption{Application-level tile arrival interval vs speed, showing an inverse relationship.}
   \label{fig:speedMotionApp}
\end{figure}

For \textit{application level} network traffic, we picked a single starting point and covered paths in the three different directions (as described in Section~\ref{sec:Data Set}), although in our data the different directions did not seem to have much impact. Each of these paths was traversed at a uniform speed, chosen among 10 different speeds in increments of five miles / hour.   The results, evident in Figure~\ref{fig:speedMotionApp}, are significantly less noisy than their device-level counterparts but continue to demonstrate the expected inverse relationship.  Moreover, since the application data comes solely from Google Maps, we can correlate spurts of traffic with tile downloads, leading to the inference of a tile size of roughly 1000 bytes. This correlation also allowed us to depict Figure~\ref{fig:speedMotionApp} in units of seconds/tile rather than seconds/byte.

\begin{figure}
\centering
   \includegraphics[scale=0.6]{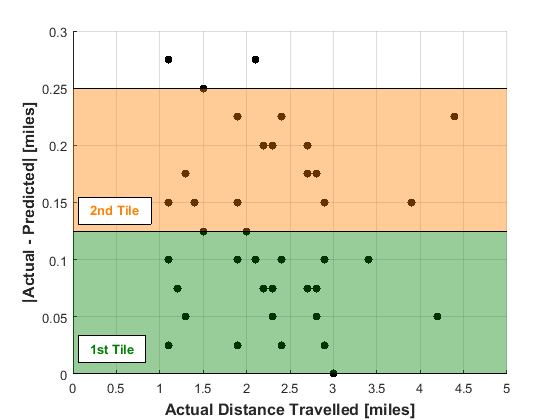}
   \caption{Actual distance traveled vs. Absolute error (difference between predicted distance and actual distance) shows accuracy within $1/4$ of a mile.}
   \label{fig:distanceMotion}
\end{figure}

Finally, we see that the relationship obtained in Figure~\ref{fig:speedMotionApp} is as follows: 
\[
	y = \frac{450}{x}
\]
In order to validate this relationship, we conducted real world driving experiments in which we drove an (unemulated) Nexus 4 phone around the city of Boston, with the default Google Maps zoom level and navigation mode \emph{not} engaged.  By correlating tile download interval and speed, we determined that each downloaded tile (at the default zoom) represents roughly a $1/8$ of a mile long by $1/8$ of mile wide patch of territory; thus, integrating tile arrivals over time allowed us to predict the overall distance traveled by the phone. We then compared our prediction to the actual distance travelled to produce Figure \ref{fig:distanceMotion}. In all, we produced $84$ recorded trips with physical distances ranging from $1$ to $5$ miles, and almost all of our predicted distances were within $\pm$ $1/4$ mile ($2$ tiles) of the actual distance, as shown in Figure \ref{fig:distanceMotion}.

As such, through unpermissioned network traffic analysis, an application developer can infer a device's speed and, by extension, the mode of transportation being used at that time.  Of course, changing the zoom level and navigation mode settings could alter the tile download statistics, and it would be an interesting extension to train our model at multiple setting parameters.

\subsubsection{Coarse Location}
\label{sec:t_loc}
Our third set of experiments attempted to correlate the network side-channel with properties of the location of the device.  To test this, we used the emulator to measure network usage as the phone moved around in different locations.  We conjecture that Google Map tiles have 
different, and potentially fingerprintable, amount of content. Intuitively, urban areas would have more information per tile, so the tiles would be denser, while those of rural areas would be sparser. Hence, we attempted to differentiate location types based on the data download rates; cities would typically have higher data download rates due to ``denser'' tiles, while rural areas would have smaller data download rates due to ``sparser'' tiles. 

We use a nearest neighbor classifier on the average download rate (in Bytes/Second) to determine whether a phone is in a rural or urban location.
We conducted this experiment using device-level traffic consisting of 432 paths (positioned in 6 urban and 6 rural areas, as discussed in Section \ref{sec:Data Set}). Across these paths, we can identify whether the device is in an urban or rural location with a Correct Classification Rate (CCR) of 67\%.  

While this is not as precise as we had hoped, this clearly shows that there is a non-zero correlation between the network side-channel and the physical location of the device.  We conjecture that with more extensive data collection and more careful modeling and analysis, much better location identification can be achieved.

\subsubsection{Path Traversal}
\label{sec:t_path}
Our final set of experiments aimed to identify a particular path of travel from a predefined set of possible paths.
Intuitively, specific paths will have a rather unique sequence of speeds along the path that may be used for classification.  Since these unique
features are hard to emulate, we have to capture actual traffic patterns and road information, these experiments were only 
performed using real-world measurements rather than emulated movement.  For this reason, we only present results for a 
limited set of paths.  However, even with this small data set, it is clear that driven paths can be predicted pretty accurately
using only the network side channel. 

We collect data on a Nexus 4 smart phone while driving the specified paths in the greater Boston area.  We recorded data usage across six different paths, with ten different recordings of each path.  
To test the correlation of path with data usage, we divide this dataset into 2 sets of paths, one labeled, and the other unlabeled. For the time series data of each path in the unlabeled dataset, we compute its Pearson Correlation Coefficient (PCC) with all paths in the labeled dataset to measure how well correlated the unlabeled path is with each labeled path.  
We then classify the unlabeled path as whichever labeled path it gets the maximum PCC with (time series originating from the same paths should be similar, and hence should have a PCC close to 1). 

We utilized the following paths, all located in the greater Boston area:
\begin{enumerate}
  \item Mountfort to Cambridge
  \item Cambridge to Mountfort
  \item General Electric (Lynn) to Charlesgate (Kenmore)
  \item YI (Synagogue - Brookline) to General Electric
  \item Mountfort to Maimo (High school in Brookline)
  \item Ivy Street to Butch (store in Brookline)
\end{enumerate} 

The results of the classification model
are presented in Table \ref{classResults}.  Here each column corresponds to a traveled path, 
while the rows indicate what path the classification outputs.  We note that some paths are 
identified quite well.  For example, paths 2 and 3 are identified correctly in over $95\%$ of cases.
Other paths are less unique, thus leading to lower classification scores.  For example, paths 1 
and 6 share a significant stretch of road, and are thus often misclassified for one another.  However,
these paths can still be distinguished from the remaining paths with high probability.  While the total
Correct Classification Rate (CCR)
across all paths is only $77\%$, this shows that some paths are uniquely identifiable while other, more similar paths,
can be grouped still providing some information about the actual location of the device.

\begin{table}[H]
\centering
\begin{center}
 \begin{tabular}{|c|c|c|c|c|c|c|}
 \hline
    \textbf{Classified as} & \multicolumn{6}{|c|}{\textbf{Actual Path}} \\ \hline
    & \textbf{Path 1} & \textbf{Path 2} & \textbf{Path 3} & \textbf{Path 4} & \textbf{Path 5} & \textbf{Path 6} \\ \hline
   \textbf{Path 1} & 60\% & 2\% & 2\% & 1\% & 10\% & 20\% \\ \hline
   \textbf{Path 2} & 0\% & 82\% & 0\% & 0\% & 16\% & 3\% \\ \hline
   \textbf{Path 3} & 0\% & 0\% & 98\% & 0\% & 0\% & 10\% \\ \hline
   \textbf{Path 4} & 0\% & 0\% & 0\% & 96\% & 10\% & 6\% \\ \hline
   \textbf{Path 5} & 0\% & 11\% & 0\% & 3\% & 64\% & 0\% \\ \hline
   \textbf{Path 6} & 40\% & 5\% & 0\% & 0\% & 0\% & 61\% \\ \hline \hline
   \multicolumn{6}{|l|}{\textbf{Total Correct Classification Rate:}} & \textbf{77\%} \\ \hline
 \end{tabular}
 \end{center}
 \caption{Path Classification Results.  The columns correspond to the actual path travelled, while the rows correspond to the identified path.}~\label{classResults}
\end{table}

\section{Attacking Web Browser Privacy} \label{sec:browse}
We next present our attacks on web browsing privacy using the network side channel.  The websites one visits may reveal private traits such as shopping interests, religious affiliation, or membership in various clubs and groups, and the ability to unmask this information and correlate it with timing, using an app which requires no-permissions, may thus pose a significant privacy concern.  Unlike the previous section, the attacks described in this section only leverage \emph{device-level} network statistics as opposed to per-app network statistics; this  information is available to all apps, without any required permissions, running on even the latest Android phones.

The rest of this section is organized as follows.  We begin with a description of the testing methodology in Section~\ref{subsec:methodology}, describing the processing and data collection for our experiments.  Then, we describe common side-channel countermeasures in Section~\ref{subsec:counter}.  Finally, we collate all our results for website classification and countermeasure robustness in Section~\ref{subsec:results}.

\subsection{Testing Methodology}
\label{subsec:methodology}
The methodology used for this attack can be divided into several sections: an Android Layer to collect network usage data on the Android device, a Preprocessing Layer to transform this raw data into feature vectors, and a Support Vector Machine (SVM) Layer to perform the website classification. 

\subsubsection{Android Layer}
\begin{figure}
	\centering
	\includegraphics[scale=0.4]{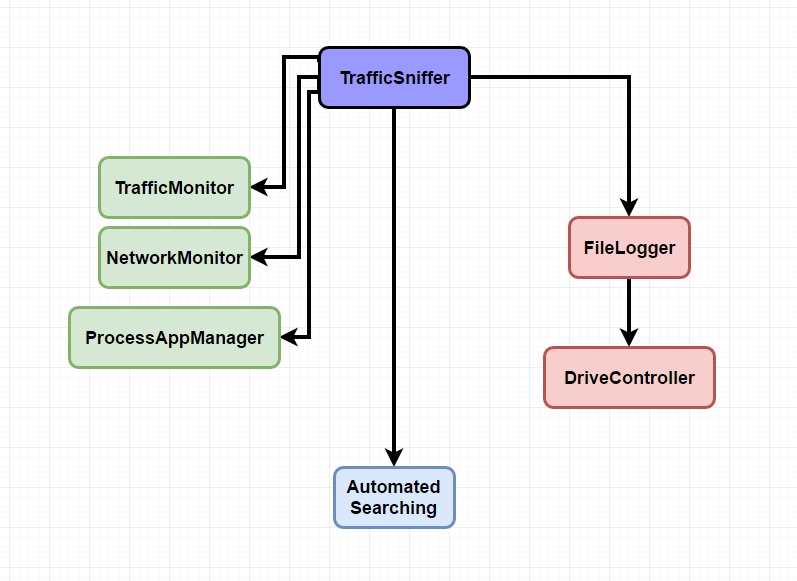}
	\caption{Android application class structure. All branches off of Traffic Sniffer run in parallel.}
	\label{fig:android}
\end{figure}

This layer consists of an Android app used to gather data, in our case the timing, size, and numbers of packets being sent or received by the device. The app itself consists of several components diagrammed in Figure \ref{fig:android}. A \texttt{TrafficSniffer} controls the application and provides a graphical interface. A \texttt{FileLogger} and \texttt{DriveController} work to upload captured data up to a Google Drive server. A \texttt{TrafficMonitor}, \texttt{NetworkMonitor}, and \texttt{ProcessAppManager} run in parallel to obtain side-channel information about the device. The \texttt{TrafficMonitor} uses the \texttt{TrafficStats} API included in the Android framework~\cite{trafficStats} to acquire packet, data, and timing information roughly every millisecond.

As mentioned in Section~\ref{sec:network}, \texttt{TrafficStats} can keep track of an individual app's network statistics using its ID. In the newer Android versions this feature has been restricted, leaving only device-level traffic statistics (with the resulting additional noise).

Test data is acquired with an \texttt{AutomatedSearching} class that performs website requests against specified URLs at thirty second intervals. The timing of the requests is recorded and then correlated to network traffic. 

Our specific experimental setup involved the sampling of 128 URLs chosen as the most popular 
sites on the Alexa top 500 most visited websites.  In addition to being very popular (and, thus, likely to be observed
in a user session), these URLs span a range of website categories (e.g., web search, news, social media, and video), ensuring
that a generally successful attack would be broad enough to cover different usage models.  In all, 19199 website requests were targeted against
these URLs, averaging roughly 150 requests per URL. All requests were performed on a Samsung Galaxy S5 device, with all standard processes running for background noise, with the sole exception of updates, which are infrequent but have exceptionally bursty utilization of network traffic.

\subsubsection{Preprocessing Layer}
Our preprocessing layer involves transforming raw data inputs into relevant feature vectors.  In order to support the variety of input data, this required experimentation to determine the optimal parameters for tasks such as frame size selection, packet filtering, and transformation of collected data into feature vectors.  In the end, we made the following design decisions.

\begin{figure}
	\centering
	\includegraphics[scale=0.5]{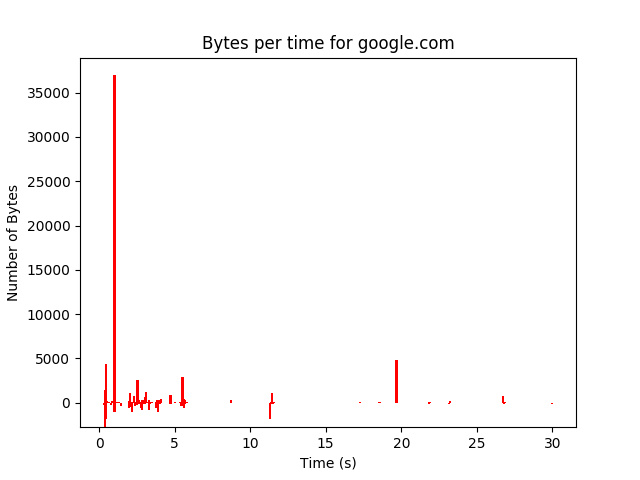}
	\caption{Bytes per second entering (positive) or leaving (negative) the phone in a single frame.}
	\label{fig:bytes}
\end{figure}

\begin{figure}
	\centering
	\includegraphics[scale=0.5]{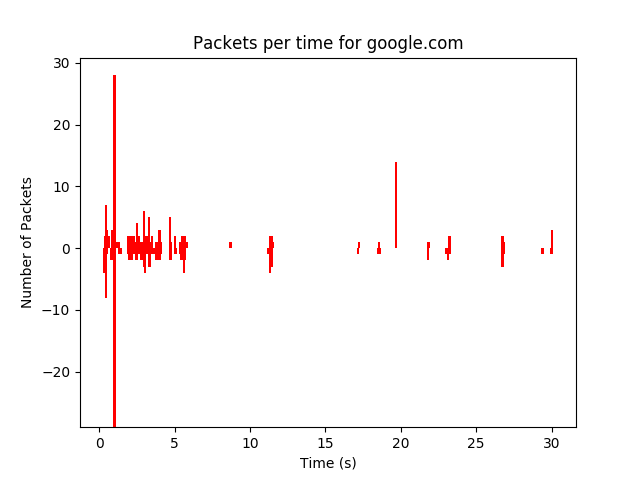}
	\caption{Packets per second entering (positive) or leaving (negative) the phone in a single frame.}
	\label{fig:packets}
\end{figure}

\paragraph{Frame Size Selection}
Our first preprocessing selection was to choose the size of an input frame from which to draw input data, or, in other words, the amount of network traffic data to correlate with a particular request.  For sake of simplicity, we used a frame size of 30 seconds, corresponding to the interval time from the \texttt{AutomatedSearching} class of the Android Layer; Figures~\ref{fig:bytes} and~\ref{fig:packets} 
show some of our captured traffic data over one such frame.  Of course, real users are not likely to methodically look at websites in 30 second intervals, and future extensions would have to take this into account.

\paragraph{Packet Filtering}
Next, we had to apply filtering to clean up noise in the collected data.  As can be seen in Figures~\ref{fig:bytes} and~\ref{fig:packets}, the ``bytes per time'' and ``packets per time'' measures are not always aligned in time, and we thus first matched a packet to a set of bytes based on their proximity to each other.  Thereafter, we filtered packets smaller than 100 bytes, which was determined experimentally, in order to reduce noise.

\paragraph{Transformation Schemes}
We also attempted to use several different transformation schemes to get feature vectors that produced good results from our classifier. Throughout these transformation schemes, we use the convention that data going into the phone is treated as a positive value, and data going out is treated as a negative value.

\begin{figure}
	\centering
	\includegraphics[scale=0.5]{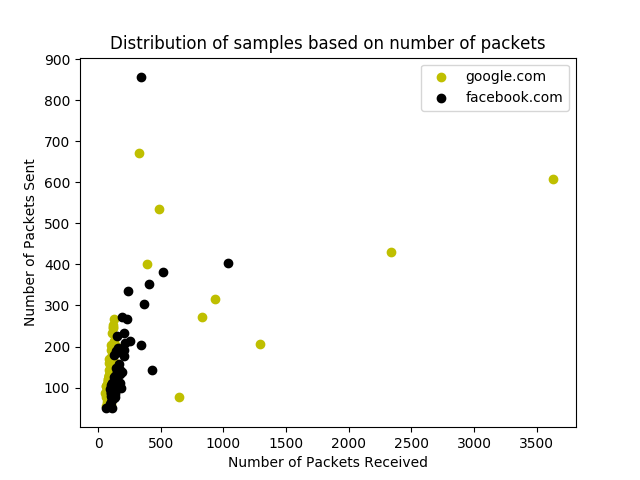}
	\caption{Representation of the data entries using the number of packets entering or leaving the phone. Yellow dots indicate website requests against \textit{google.com}, and black dots indicate requests against \textit{facebook.com}.}
	\label{fig:datamap}
\end{figure}

\begin{itemize}
	\item \textbf{``onion'' - }This transformation scheme uses the features presented by~\cite{onion}, derived from raw data, which include: aggregate byte markers for data direction changes rounded to 600, aggregate packet markers for data direction changes, total transmitted and received bytes rounded to 10000, total transmitted and received packets rounded to 15, different occurring packet sizes, and percentage of incoming packets versus outgoing packets rounded to steps of 5. The rounding boundaries were selected from the paper. The values are then aggregated into a single feature vector, and, together, give us a statistical view of the data while reducing noise through rounding.
	\item \textbf{term frequency transformation with cosine normalization - }Following the work in~\cite{multi_bayes}, we generated a term frequency vector whose $i$-th component corresponds to the frequency $f_{x_i}$ of packets of size $x_i$.  The term frequency transformation is then given by $$f^*_{x_i} = \log(1+f_{x_i}),$$ and its cosine normalization is determined by dividing by the Euclidean length ($|| \cdot ||$) of the transformed vectors: $$f_{x_i}^{norm} = \frac{f^*_{x_i}}{||(f^*_{x_1}, ..., f^*_{x_n})||}.$$
	\item \textbf{number of packets - } This transform simply takes the number of incoming packets and the number of outgoing packets per frame, and combines them into a two dimensional array, as in Figure \ref{fig:datamap}.
\end{itemize}
In our testing with different parameters, we found ``number of packets'' to be both the simplest and the most effective transformation.

\subsubsection{SVM Layer}
To assign a URL to an input feature vector, we turn to a machine learning algorithm called a Support Vector Machine, or SVM, which are used for binary classification problems \cite{ML_book}. The primary goal of an SVM is to create a decision boundary between two sets of data, where, ideally, all data points on one side of a boundary have the same label. SVMs further define a set of margins around the decision boundary, so that no data points exist between a margin and a decision boundary.  By creating a margin around the decision boundary, the SVM reduces any overfitting that could be present in a more basic perceptron. Instead of choosing the simplest boundary that fits the data, SVMs create a more distinct separation between classes, which may allow new, testing data to be classified more accurately.

If the data an SVM is trying to classify is not linearly separable, other functions may be used to calculate the decision boundary and margins. In our case, we used the radial basis function (RBF) to calculate the decision boundary. Since our algorithm works on a linear learning scheme, it can be difficult to create a proper non-linear decision boundary, and we thus make use of the "kernel trick" or "kernel method" \cite{ML_book} to map the radial function inputs into a different dimensional space, where the calculations become straightforward to perform. The mapping function is referred to as a kernel, and for RBF it is:
$$ K(x,x') = e^{- \gamma ||x-x''||^2}, $$
where parameter $\gamma$ is described below, $x$ and $x'$ are feature vectors, and $|| \cdot ||$ represents the Euclidean distance between the vectors.  For a given set of features, these mappings can be put together into a kernel (or Gram) matrix $K$, whose entries $K_{ij} = K(x_i,x_j)$.

There are two parameters we have to balance to obtain an optimal classification: the $\gamma$ parameter, which determines the smoothness of the decision boundary, and cost parameter (C) that determines the cost of having an imperfect boundary (i.e., how bad is it if a point is misclassified).

We trained the classifier by first calculating the kernel matrix as described above, and feeding the kernel matrix and the feature vectors into a quadratic programming optimization. 
From there, testing inputs can be passed into the model, which will return a value that determines what side of the boundary that input lies on, giving its label.
Our code for this is all written in python and makes extensive use of \texttt{numpy}, for numerical support, and \texttt{cvxopt}, for convex optimization.

We use SVMs to perform three different classifications.

\paragraph{One vs. One Classification}
First, we evaluated how well our classifier can distinguish which of two URLs has been visited.
The test for a one on one comparison is fairly straightforward. We use cross-validation to separate the input data into five sections.  Each section takes a turn being the testing section, while the remaining four sections are used for training. We then average the correct classification rate to get the final rate for the test. This is then repeated across all URL pairs giving us a perspective on how URLs match up. The results in this classification are used to determine how to create different matches of more than two URLs. 

\paragraph{One vs. All Classification}
Next, we designed a classifier to identify URLs that are sufficiently unique in their bandwidth usage that they can be distinguished from all the rest.
The test for a ``one vs. all'' classification is similar to that of ``one vs one'', except that one of the labels being trained now has a set of 127 URLs, allowing us to treat the 127 URLs as a single category. By performing this test we can determine which URLs are separable from the rest.

\paragraph{Website Identification}
Finally, we investigated the feasibility of uniquely identifying the URL of a visited website among 128 possibilities.  We consider two different techniques for doing such identification

\paragraph{Cascade:}
Because the SVM we are using can only perform a binary classification, we need to expand the number of test we do to determine the URL. We do this by separating all 128 into two groups of 64 vs. 64. The sample is classified in one group or the other. We then split that group into 32 vs. 32, and continue classifying, eventually reaching the ``one vs. one'' level and determine the final URL. To determine which URLs are in which group, we used both a random selection method, and a greedy separation The random selection simply randomly allocated equal numbers of URLs to each side. The greedy separation randomly selected the first URL, then used the highest ``one vs. one'' comparison result with that URL to select a second URL. It continued this selection for each entry in the current pool.

\paragraph{One vs. One Tree:}
An alternative method to determining which URL a test sample belongs to is performing a series of ``one vs. one'' matches and obtaining a list of resulting URLs then repeating. This works seperating out the 128 URLs into 64 groups of two. A SVM is trained for each of these groups, and then a test sample is fed through the models. The models will select one of the two URLs they are trainged for. These URLs that are selected are then taken up to the next level, where they are seperated into 32 groups of two. This repeats until we classify the final two URLs. The selection for which two URLs to compare was performed both randomly and greedily.

\subsection{Countermeasures}
\label{subsec:counter}
Next, we highlight several countermeasures that are often proposed against traffic analysis problems. We will describe these countermeasures here, then show how effective they are against our model.

\begin{itemize}
	\item \textbf{Random-session padding} - adding random amounts of padding to different sessions, causing some requests to give a different byte profile than others for the same URL.
	\item \textbf{Random-packet padding} -  adding a random amount of byte padding to each packet, causing sizes to fluctuate significantly.
	\item \textbf{Pad to Ceilings} - padding each packet until they all appear to be the same size.
	\item \textbf{Exponential Padding} - random padding on a large scale, increasing the sizes of packets by a factor of 2.
	\item \textbf{Linear Padding} -  padding each packet by the same amount.
	\item \textbf{Pad to max} - padding the packets to the maximum allowed amounts.
	\item \textbf{Random Packet pad to max} - padding random packets to their maximum amount, leaving others untouched.
	\item \textbf{Random Packet insertions} - adding random packets of random size to the data stream.
	\item \textbf{Uniform Adding} - confusing the packet data stream in a more uniform manner.
	\item \textbf{Session Random Adding} - adding random amounts of packets to random sessions, leaving other sessions unchanged.
	\item \textbf{Proxies} - using a proxy to aggregate traffic, a common suggestion against traffic analysis attacks; the methods presented will fairly easily overcome this defense, although with the potential addition of noise, potentially harming classification.
\end{itemize}

\subsection{Browsing Results}
\label{subsec:results}
We present the results of three different sets of experiments.  The first set of experiments is used to determine the optimal parameters and transformation to use in the preprocessing phase.  The next set of experiments evaluates the performance of the different classification tasks after this preprocessing is performed.  Finally, the third set of experiments examines the effects of the proposed countermeasures.

\begin{figure}[t]
	\centering
  \includegraphics[scale=0.5]{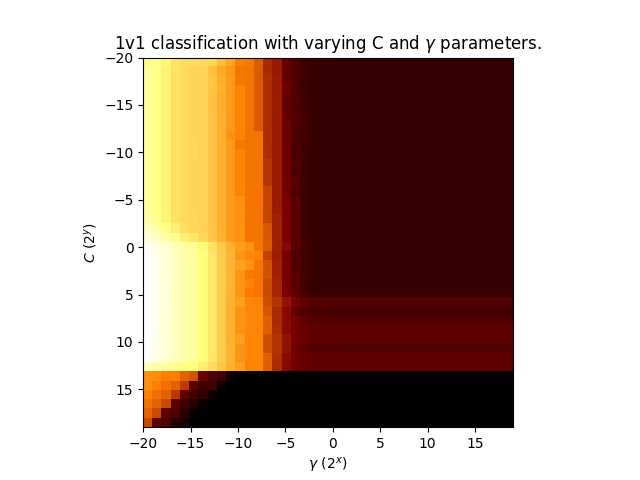}
	\caption{Correct Classification Rate (CCR) when using different C and $\gamma$ parameters. In this scale, lighter indicates better CCR. Values of C and $\gamma$ are 2 raised to the power of any given tick mark.}
	\label{fig:c_and_gamma}
\end{figure}

The Correct Classification Rate (CCR) is used here to show the ratio of correctly classified samples to all samples:

$$ CCR = \frac{Y_{correct}}{Y_{total}}.$$

In our graphs, we multiply this ratio by 100 to get a percentage value for ease of reading.

\begin{figure}
	\centering
	\includegraphics[scale=0.5]{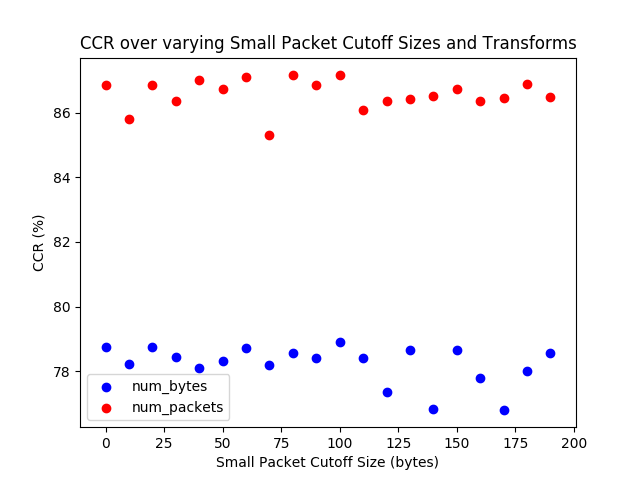}
	\caption{Correct Classification Rate (CCR) when using different cutoff filter sizes and transforms.}
	\label{fig:spf_transform}
\end{figure}

\subsubsection{Choosing Parameters}
After running many tests against different parameters we see the optimal classification parameters are \texttt{filter size} = 100 bytes, $C$ = 256, $\gamma$ = .000000954, using the ``number of packets'' transformation. These result indicate that coarser separation produces better classification. The high filter size indicates that small packets are fairly common across different websites, and thus not of particular use in differentiation. Having a very high $C$ coefficient and a very low $\gamma$ parameter indicates a non-smooth boundary and small margins, which may cause some overfitting. Figure~\ref{fig:c_and_gamma} shows the relationship of C and $\gamma$ with the correct classification rate, while Figure~\ref{fig:spf_transform} shows the effect of different cutoff filters and transforms.

An interesting result is that despite the availability of more complex processing strategies, it is the most simple method, looking at the number of packets, that proved most effective. This is likely due to the low resolution of our data; previous work \cite{onion}, \cite{multi_bayes} was able to use very fine-grained packet sizing and timing, making complex methods much more effective. However, when there is uncertainty in the timing and size, the simpler methods tend to avoid the over-fitting that plague more complex methods.

\begin{table}
	\begin{center}
		\resizebox{\columnwidth}{!}{%
			\begin{tabular}{|c|c|c|c|c|c|c|c|}
				\hline
				& \textbf{Dailymail} & \textbf{Samsung} & \textbf{Yahoo} & \textbf{Amazon} & \textbf{Google} & \textbf{Twitch} & \textbf{Tumblr}\\
				\hline
				\textbf{Alibaba} & 91.29\% & 73.75\% & 94.48\% & 77.21\% & 90.60\% & 69.95\% & 84.07\%\\
				\hline
				\textbf{Dailymail} & & 90.68\% & 84.11\% & 88.93\% & 94.82\% & 93.39\% & 85.73\%\\
				\hline
				\textbf{Samsung} & & & 91.97\% & 84.39\% & 93.89\% & 86.60\% & 60.04\%\\
				\hline
				\textbf{Yahoo} & & & & 94.48\% & 94.37\% & 96.07\% & 93.04\%\\
				\hline
				\textbf{Amazon} & & & & & 95.12\% & 90.08\% & 87.35\%\\
				\hline
				\textbf{Google} & & & & & & 87.21\% & 95.04\%\\
				\hline
				\textbf{Twitch} & & & & & & & 91.15\%\\
				\hline
				\hline
			\end{tabular}%
		}
		\caption{8x8 Sample of 1 vs 1 test results.  The values in the table show the CCR when comparing the two sites given by the row and column labels.}
		\label{1v1_test_result_sample}
	\end{center}
\end{table}

\begin{figure}
	\centering
  \includegraphics[scale=0.5]{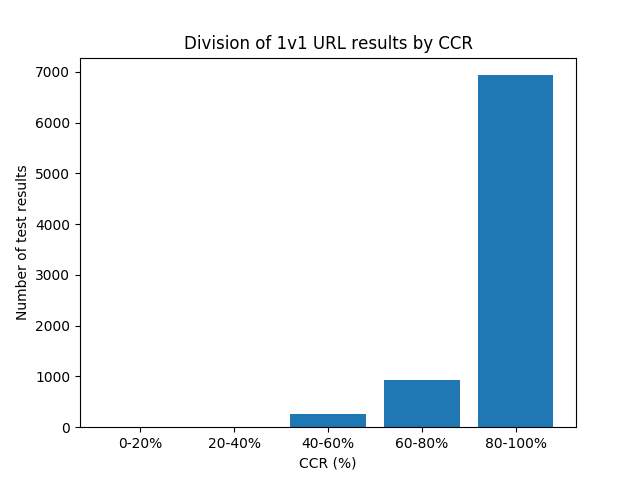}
	\caption{Distribution of correct classification rates for 1 vs 1 test.}
	\label{fig:1v1_results}
\end{figure}

\subsubsection{Classification Performance}
\paragraph{One vs. One Classification}
The "one vs. one" classification test was run against all pairs of URLs to give us a view of how different URLs compared with each other. This gives us a basis to form classification groups in later tests. It also gives us a view of which URLs are very distinct from each other, and which are similar. Overall, our average classification rate was 87.88\%. Our best classification was 97.42\% between \textit{google.com} and \textit{t.co}. Table~\ref{1v1_test_result_sample} shows a sample of the test results, and Figure~\ref{fig:1v1_results} shows the overall distribution. It is clear that it is easy to tell apart most URLs on a one on one basis.

However, some of the matches produce very low results, from which we can deduce those websites appear to be similar. Some of these websites makes sense: different versions of Google home page, \textit{google.com} vs \textit{live.com}. Others are more surprising: \textit{google.com} vs \textit{instagram.com}. Although the results found here may be circumstantial, further study might unearth a relationship between a low classification rate and a website type.

\begin{figure}
	\centering
	\includegraphics[scale=0.5]{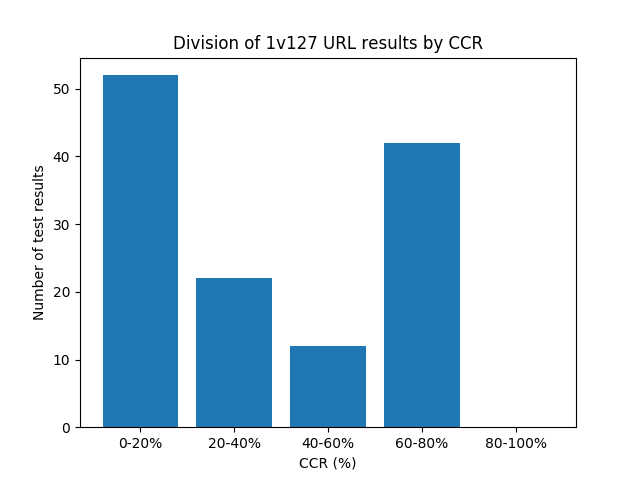}
	\caption{Shows distribution of classification rates for 1 vs all test.}
	\label{fig:1vall_results}
\end{figure}

\paragraph{One vs. All Classification}
The ``one vs. all'' classification test was run against all URLs to see which could be determined easily. Figure~\ref{fig:1vall_results} shows the distribution of this classification. Most results are understandably low. However, for a set of 35 URLs (e.g., facebook, baidu, dropbox), our CCR is around 0.75. The reason behind this result is that when we train our SVM to classify 127 different URLs as one class, we are essetially creating an "average" website to classify. Websites that do not fit this "average" become easy to spot, allowing fast identification. Although this comparison will not yield good results for all URLs, an attacker can use it to quickly identify usage of specific websites. In addition, along with the findings of the ``one vs. one'' test, this could be taken further to establish a relationship between website types.

\begin{figure}[p!]
	\centering
	\includegraphics[scale=0.5]{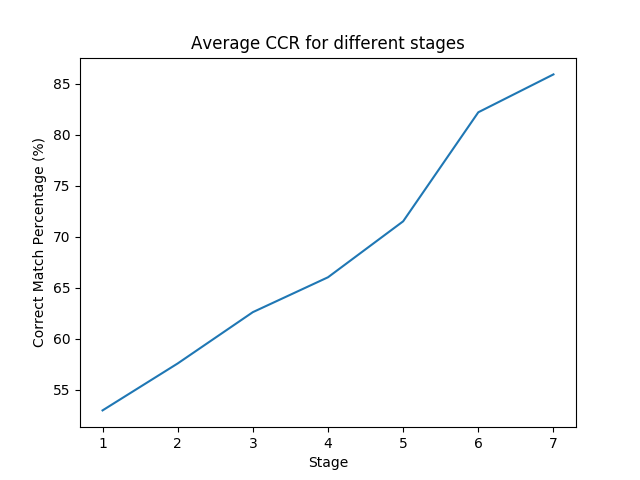}
	\caption{Average CCR for each stages of the cascade. This shows how each individual stage performed, regardless of how the previous stage classified its data. The x-axis indicates the number of URLs in the stage.}
	\label{fig:cascade_size}
\end{figure}

\begin{figure}[p!]
	\centering
	\includegraphics[scale=0.5]{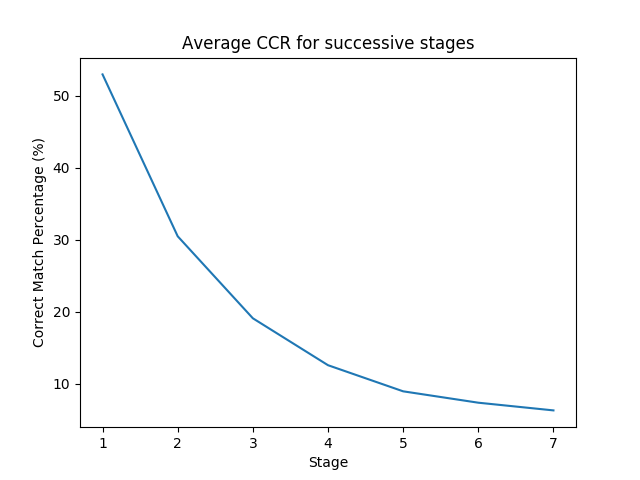}
	\caption{Average successive CCR for each stage of the cascade. This shows how each stage performed, given the previous stage's results. The x-axis indicates the number of URLs in the stage. The final x-tick indicates the final classification of 1 URL against the rest.}
	\label{fig:cascade_cascade}
\end{figure}

\begin{figure}[p!]
	\centering
	\includegraphics[scale=0.5]{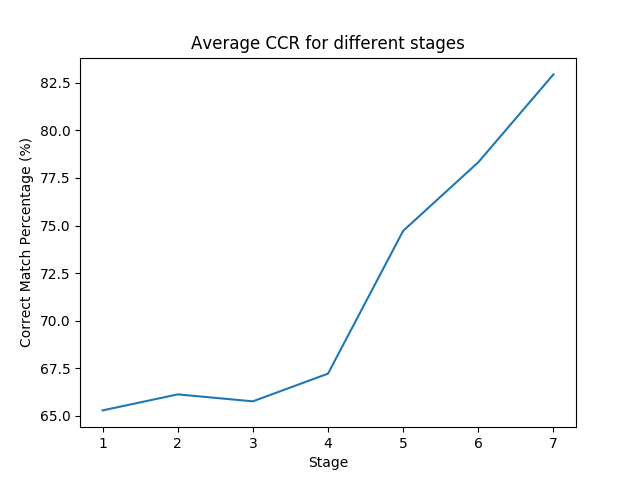}
	\caption{Average CCR for each stage of the cascade using greedy selection. This shows how each individual stage performed, regardless of how the previous stage classified its data. }
	\label{fig:cascade_size_greedy}
\end{figure}

\begin{figure}[p!]
	\centering
	\includegraphics[scale=0.5]{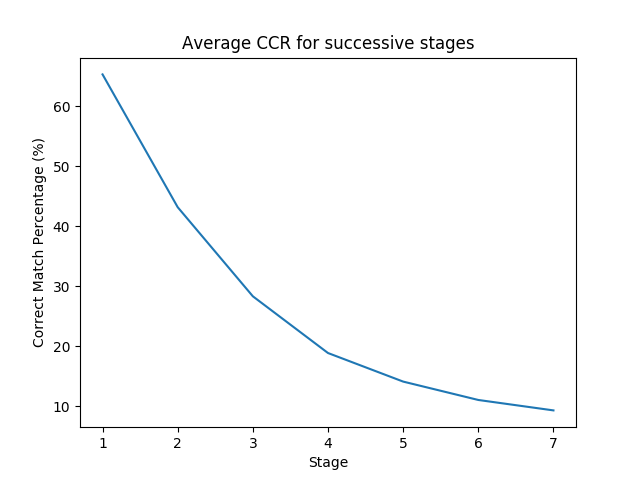}
	\caption{Average successive CCR for each stage of the cascade using greedy selection. This shows how each stage performed, given the previous stage's results. }
	\label{fig:cascade_cascade_greedy}
\end{figure}

\begin{figure}
	\centering
	\includegraphics[scale=0.5]{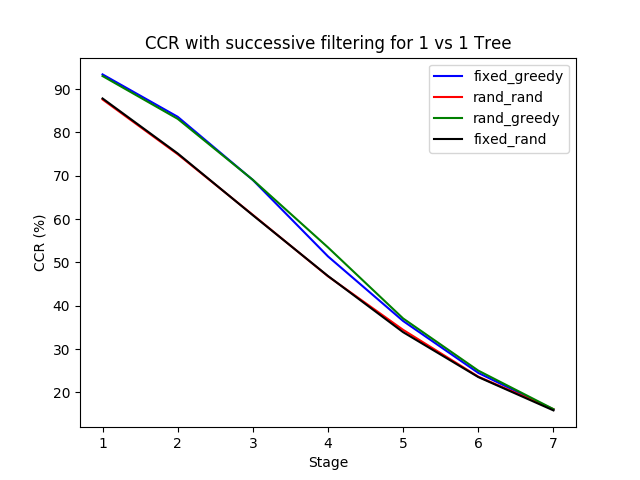}
	\caption{Average successive CCR at each stage of the tree. Initially starting with all URLs paired against each other, we successively eliminated less likely URLs to get a final result.}
	\label{fig:complete_grid}
\end{figure}

\paragraph{Website Classification}
\paragraph{Cascade}
This test was done in two parts. The difference between the parts] was how the selection of URLs for each cascade was done. In the first part, random selection was done. In the second part, URLs were chosen by a greedy approach by using the highest ``one vs. one'' classifications. 

Figure~\ref{fig:cascade_size} shows the correct classification rate at each stage of the first part. Each stage represents the next level of the cascade. For example, the first stage compares 64 URLs against the other 64 URLs. The second stage compare 32 URLs against the other 32 URLs as passed down from one of the 64 URL groups from the stage above. In this graph we are showing how the stage performed individually, that is, if given a set of test samples, how well can this stage separate the samples into the correct groups.

As can be seen, the fewer URLs at a given stage, the better the classification rate. This indicates it is easier to differentiate URLs individually than in aggregate. 

Meanwhile, Figure~\ref{fig:cascade_cascade} shows the result as the stages work sequentially. This means we first run our test samples through the first stage. Those that were classified correctly are then run through the second stage. We repeat this process until we classify each test sample into a given URL. As can be seen, the further stage we get to, the lower our CCR becomes as classification errors in previous stages become compounded in the later ones. 

Our final CCR is 6.32\%, which is clearly better than the random guessing situation (with a CCR of 0.78\%). However, we are not able to get as high results as some previous studies \cite{peekaboo},~\cite{onion}, and~\cite{data_usage_statistics}. This in part is due to the classification method and filters, but is probably in larger part related to the limited information available in this attack method.

Figure~\ref{fig:cascade_size_greedy} shows our attempt to improve classification using a greedy selection method. This gives us modest gains at each of the stages. Figure~\ref{fig:cascade_cascade_greedy} shows the result of the cascade, which has a final CCR of 9.34\%. We can learn from this that the method of separating which URLs are compared can greatly increase our accuracy.

\paragraph{One vs. One Tree}

Finally, we changed the approach in the cascade method to see if using a single classification improved our results. Figure~\ref{fig:complete_grid} shows the progression through each stage of the tree. For this method we also tested different selection approaches by determining if the URLs compared in each trial were random, fixed, or greedily chosen. Our best result was with the first URL being randomly selected, and the second being greedily extracted. We obtained a CCR of 16.14\%. 

Overall, we can see that when we have a large number of URLs to select from, the noise and limited information in our data starts to take effect by greatly reducing the final classification rate. However, even with this limitation, an attacker can reliably deduce a small subset of URLs a user can be visiting. Further, with greater variability in matchup selection, even higher rates could be achieved.	

\subsubsection{Countermeasure Effectiveness}
Finally, we performed tests to determine the effectiveness of standard side-channel countermeasures. The results for this can be seen in Table~\ref{countermeasures}. These tests were done using the same parameters as the binary classification of \textit{google.com} vs. \textit{facebook.com}.
\begin{table}
	\begin{center}
			\begin{tabular}{|c|c|}
				\hline
				\textbf{Countermeasure} & \textbf{Correct Classification Rate} \\
				\hline
				\hline
				\textbf{Random-session padding} & 88.13\%\\
				\hline
				\textbf{Random-packet padding} & 92.56\% \\
				\hline 
				\textbf{Linear padding} & 87.16\% \\
				\hline
				\textbf{Exponential Padding} & 91.66\%\\
				\hline
				\textbf{Pad to ceilings} & 87.16\%\\
				\hline
				\textbf{Pad to max} & 93.13\% \\
				\hline
				\textbf{Random-packet pad to max} & 85.27\% \\
				\hline
				\textbf{Random Packet Insertions} & 50.00\% \\
				\hline
				\textbf{Uniform adding} & 56.12\% \\
				\hline
				\textbf{Session-random} & 50.00\%\\
				\hline
				\hline
			\end{tabular}%
		\caption{Correct classification rate after different countermeasures when applied to Google vs Facebook binary classification.}
		\label{countermeasures}
	\end{center}
\end{table}	
By looking at the countermeasures, we see that ``byte padding'' is not very effective regardless of what scheme is used. Although there is a slight decrease in accuracy, likely due to interference from using the filter cutoff, the decrease in classification is not significant. On the other hand, introducing random packets into the data-flow causes a severe decrease in correct classification. This is likely due to the transformation at work. Since we are looking at the number of packets being used, changing that by a random value causes significant noise in the input samples.

The other countermeasure commonly suggested against traffic analysis attacks is the use of proxies, a sample of whose effect we already see by aggregating all network traffic on the device.  The increased noise of a large proxy might be mitigated by more precise measurement techniques.


\section{Conclusions}  \label{sec:conclusion}
We have demonstrated the potential for significant user privacy violations on an Android phone from the network side-channel.  Our specific attack scenarios focused on
identifying a user's location and browsing patterns, which may be correlated with deeply personal user information.  More concretely, our results show that a variety of sensitive information about a user's location and web browsing habits are leaked from the networking metadata available to \emph{any} application on Android phones.

Rather than being an esoteric technical vulnerability, we demonstrate the effectiveness of these attacks in a variety of experimental setups.  The type of information leaked through these channels might be used to infer very personal information about the user, such as religious/political affinities and medical conditions.

Ultimately, our results illustrate the need to better understand and regulate the metadata that is made available during routine operations of a mobile device.  It is unreasonable to expect users, and even most app developers, to be aware of the all such side channels and their implications on the privacy of their data.  Thus, we cannot expect them to properly navigate this challenging space when choosing what data to share or to collect with their apps.  Instead, we hope that by demonstrating such attacks we will help guide design and regulatory decisions made by industry and regulatory bodies to put in place protections that will steer app developers and users clear from such vulnerabilities.  We note that Google's recent decision to downgrade some of the functionality of the \texttt{TrafficStats} class is a step in this direction as it significantly reduces the amount of freely available network side-channel information, and we hope that our work will lead to further strengthening of user privacy.  
%
%
%
\bibliographystyle{splncs04}
\bibliography{refs}

\end{document}